\documentclass[a4paper]{article}
\usepackage{epsfig,amsmath,amsfonts,amssymb,setspace,multirow,textcomp}
\usepackage[T1]{fontenc}
\makeatletter
\renewcommand{\@oddhead}{\textit{Advances in Astronomy and Space Physics} \hfil}
\renewcommand{\@evenfoot}{\hfil \thepage \hfil}
\renewcommand{\@oddfoot}{\hfil \thepage \hfil}
\makeatother

\renewenvironment{thebibliography}[1]{\begin{oldthebibliography}{#1}\setlength{\parskip}{0ex}\setlength{\itemsep}{0ex}}{\end{oldthebibliography}}

\newcommand{\folder}{pic-col}
\usepackage{enumerate}
\usepackage{caption}
\usepackage{cases}
\usepackage{multicol}
\usepackage{cite}
\usepackage{hyperref}
\usepackage{ulem}
\hypersetup{colorlinks = true, citecolor = {blue}, urlcolor = {blue}, linkcolor={blue}}
\renewcommand{\arraystretch}{1.2}
\DeclareMathOperator\erf{erf}

\begin{document}
\fontsize{11}{11}\selectfont 
\title{Kinematic characteristics of the Milky Way globular clusters \\
based on \textit{Gaia} DR2 data}
\author{\textsl{I.\,V.~Chemerynska$^{1,7}$\footnote{{\href{mailto:chemerynskaira@gmail.com}{\tt chemerynskaira@gmail.com}}}, 
M.\,V.~Ishchenko$^{2,3}$, 
M.\,O.~Sobolenko$^2$, 
S.\,A.~Khoperskov$^{4,5}$, 
P.\,P.~Berczik$^{7,4,8,3}$}}
\date{\vspace*{-6ex}}
\maketitle
\begin{center}{\small 
$^{1}$Taras Shevchenko National University of Kyiv, Glushkova ave., 4, 03127, Kyiv, Ukraine\\
$^{2}$Main Astronomical Observatory, National Academy of Sciences of Ukraine,\\
27 Akademika Zabolotnoho St., 03143, Kyiv, Ukraine\\
$^3$Fesenkov Astrophysical Institute, Observatory 23, 050020 Almaty, Kazakhstan\\ 
$^4$Leibniz-Institut für Astrophysik Potsdam (AIP), An der Sternwarte 16, 14482 Potsdam, Germany\\ 
$^5$GEPI, Observatoire de Paris, Université PSL, CNRS, 5 Place Jules Janssen, 92190 Meudon, France\\
$^6$National Astronomical Observatories and Key Laboratory of Computational Astrophysics, Chinese Academy of Sciences,\\
20A Datun Rd., Chaoyang District, 100101 Beijing, China\\
$^7$Institut d’Astrophysique de Paris, CNRS - Sorbonne Universite, 98bis boulevard Arago, 75014 Paris, France\\
$^8$Konkoly Observatory, Research Centre for Astronomy and Earth Sciences, Eötvös Loránd Research Network (ELKH),
MTA Centre of Excellence, Konkoly Thege Miklós út 15-17, 1121 Budapest, Hungary\\}
\end{center}

\begin{abstract}
Using the data from \textit{Gaia} (ESA) Data Release~2 we performed the orbital calculations of globular clusters~(GCs) of the Milky Way. To explore possible close encounters~(or collisions) between the GCs, using our own developed high-order $\varphi$-GRAPE code, we integrated backward and forward orbits of 119 objects with reliable positions and proper motions. In calculations, we adopted a realistic axisymmetric Galactic potential~(\textit{bulge + disk + halo}). Using different impact conditions, we found four pairs of the six GCs that may have experienced an encounter within twice the sum of the half-mass radii (``collisions'')
over the last $5$~Gyr: Terzan~3 -- NGC~6553, Terzan~3 -- NGC~6218, Liller~1 -- NGC~6522 and Djorg~2 -- NGC~6553. \\[1ex]

\end{abstract}

\begin{multicols}{2}
\section*{\sc introduction}
\indent \indent 
It is believed that GCs in the Milky Way~(MW) are old gravitationally bound systems of stars with typical ages $\gtrsim6$~Gyr and masses $\gtrsim10^{4}\rm M_{\odot}$~\cite{Kharchenko2013}. These objects are a powerful tool to examine the Galactic structure and assembly history at different scales from the star clusters formation to hierarchical merger events~\cite{Kruijssen2020}. The recent precise astrometric measurements from \textit{Gaia} Data Release~2~(DR2)~\cite{Gaia2018} provide a possibility to measure the mean proper motions for $\approx 150$~GCs in the  MW which makes it possible to study the orbital evolution of the GCs system as the whole. 

In this work, we aim to explore the close encounters between different GCs and find the pairs of the GCs which 
have an encounter within twice the sum of the half-mass radii (``collisions'').
In order to do that, using two GCs catalogs~\cite{Baumgardt2019,Vasiliev2019}, we study the dynamics of the GCs as the test-particles in the axisymmetric MW-like potential over the last $5$~Gyr~\cite{Gnedin1997, Pichardo2004, Allen2006, Allen2008, Moreno2014, Perez2018, Perez2020}.  

\begin{figure*}[htbp!]
\centering
\includegraphics[width=0.31\linewidth]{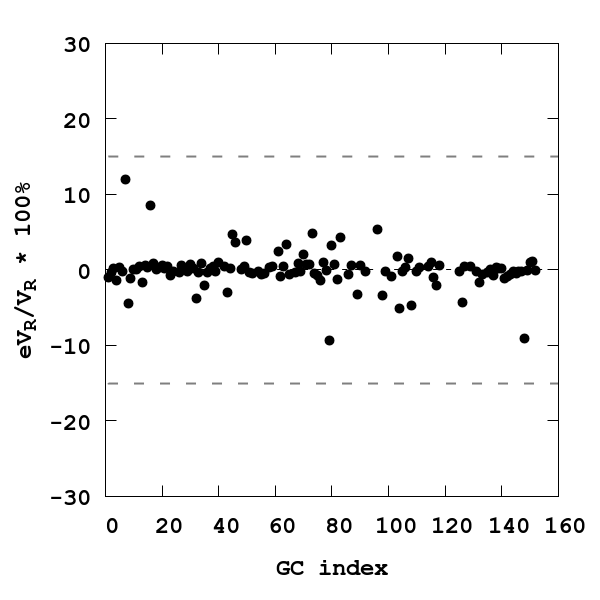} 
\includegraphics[width=0.31\linewidth]{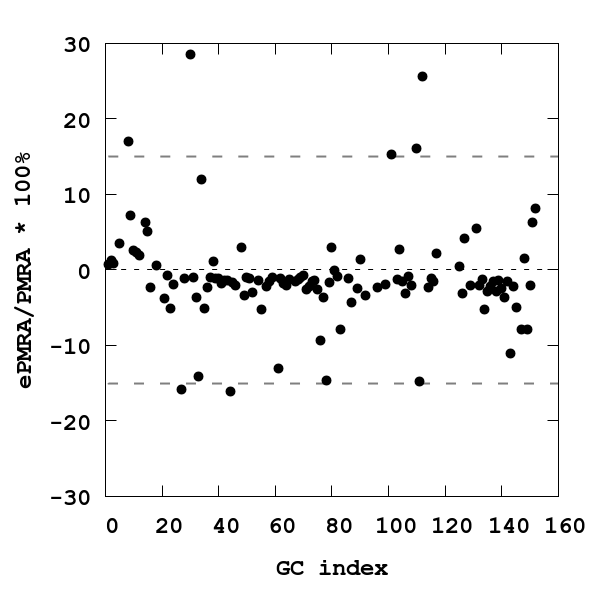} 
\includegraphics[width=0.31\linewidth]{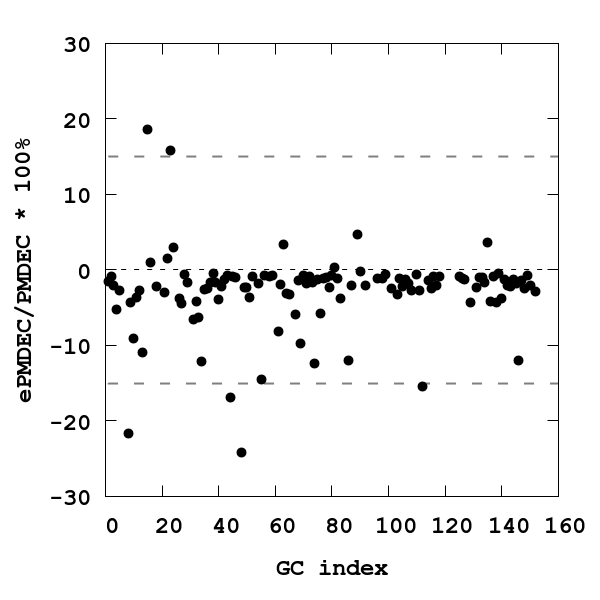} 
\caption{Distribution of the GCs measurement errors for radial velocity $v_{r}$~(left) and proper motions in right ascension~($\mu_{\alpha\ast}$, center) and in declination~($\mu_{\delta}$, right). Dashed gray horizontal lines indicate 15\% confidence range.}
\label{fig:PM}
\end{figure*}

\section*{\sc globular cluster sample}
\indent\indent 

Prior to the orbital integration, we prepared a complete catalog of the GCs. For this purpose, we combined two recent catalogs~\cite{Vasiliev2019,Baumgardt2019} which contain information on $152$ objects~(see Table~\ref{tab:GC3}). The resulting catalog contains the complete phase-space information required to set the initial conditions in our simulations: right ascension~(RA), declination~(DEC) and distance~(D), proper motions $\mu_{\alpha\ast}=\mu_{\alpha}\cos\delta$,  $\mu_{\delta}$ and radial velocity $v_{r}$.

To avoid the calculation of the GCs orbits with large  uncertainties of initial conditions we first performed an analysis of the errors of the \textit{Gaia} measurement required for our simulations. In Fig.~\ref{fig:PM} we show the relative errors for the radial velocity and proper motions where each GC has its own index~(see Table~\ref{tab:GC3}). The obtained uncertainties for the radial velocity ($v_{r}$) lie in the range of [-10\%, +15\%] for all studied GCs. The analysis of the uncertainties for proper motions ($\mu_{\alpha\ast}$, $\mu_{\delta}$) showed that its value can exceeds 30\% for some individual GCs in our sample. Therefore, the GCs with the relative error larger than $30\%$ for radial velocity and proper motions were removed from our catalogue and further analysis. See 8 GCs marked as \textit{me}~(measurement error) in Table~\ref{tab:GC3}.

For calculating positions and velocities in the Galactocentric rest-frame (for basic coordinate transformation see \cite{JS1987}), we accepted an in-plane distance of the Sun from the Galactic center and the plane as 
$X_{_{\odot}} = 8.178$~kpc~\cite{Gravity2019} and $Z_{_{\odot}}=20.8$~pc~\cite{Bennett2019}, a velocity of the Local Standard of Rest~(LSR), $V_{\rm LSR}=234.737$~\cite{Mard2020}, and a peculiar velocity of the Sun with respect to the LSR, $U_{_{\odot}} = 11.1$~km~s$^{-1}$, $V_{_{\odot}} =12.24$~km~s$^{-1}$, $W_{_{\odot}}=7.25$~km~s$^{-1}$~\cite{Schonrich2010}.

We assume the initial positions and velocities of the GCs in the heliocentric coordinate system as $(X,Y,Z)$ and $(U,V,W)$, respectively. As a result, the initial positions $(x,y,z)$ and velocities $(u,v,w)$ of the GCs in the rectangular galactic coordinates can be derived from the positions and velocities of the GCs in the heliocentric coordinate system $(X,Y,Z)$ and $(U,V,W)$ as follows:
\begin{equation}
\begin{cases}
x = X + X_{_{\odot}} + X_{_{\rm LSR}},\\
y = Y + Y_{_{\odot}} + Y_{_{\rm LSR}},\\
z = Z + Z_{_{\odot}} + Z_{_{\rm LSR}},
\end{cases}
\end{equation}
\begin{equation}
\begin{cases}
u = U + U_{_{\odot}} + U_{_{\rm LSR}},\\
v = V + V_{_{\odot}} + V_{_{\rm LSR}},\\
w = W + W_{_{\odot}} + W_{_{\rm LSR}}\,,
\end{cases}
\end{equation}
where we adopt $U_{_{\rm LSR}}=W_{_{\rm LSR}}=0$ and $Y_{_{\odot}}=0$.

\begin{figure*}[htbp!]
\centering
\includegraphics[width=0.45\linewidth]{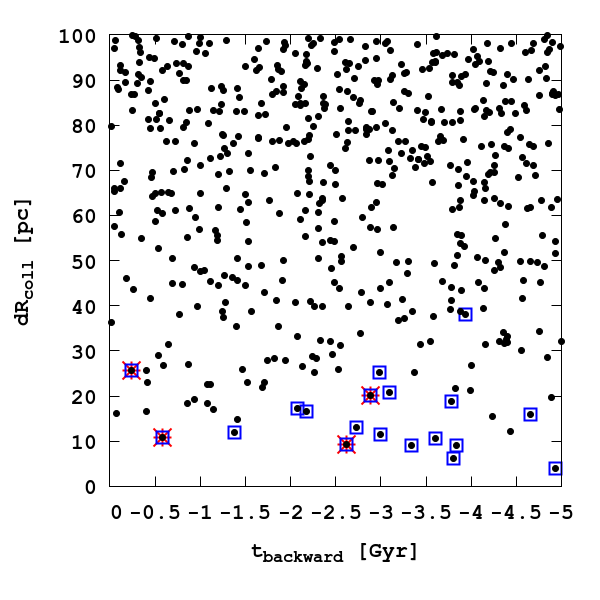}
\includegraphics[width=0.45\linewidth]{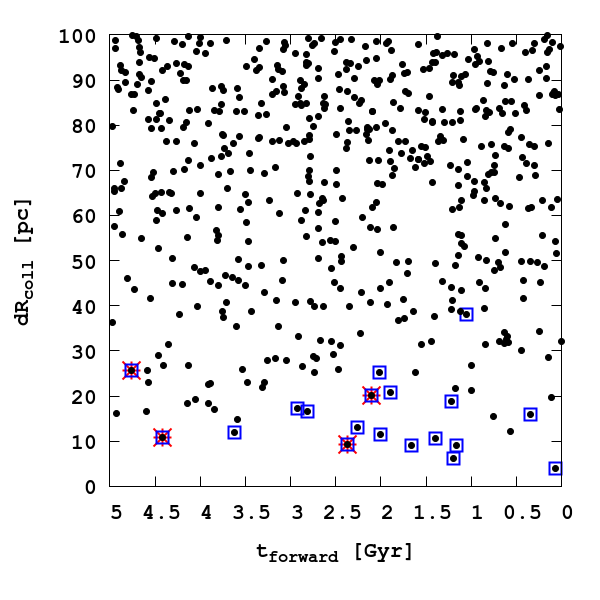}
\caption{Relative separation of the GC collision pairs~(black dots) in backward~(left) and forward~(right) integration. Open squares indicate the collisions with $dR_{\rm coll}<2(R_{{\rm hm},i}+R_{{\rm hm},j})$ and 
asterisks indicate the collisions with $dV_{\rm coll}<200$~km~s$^{-1}$.}
\label{fig:allcoll}
\end{figure*}

\section*{\sc orbits integration}

\indent \indent 
For the GCs orbit integration we adopted the MW-type gravitational potential based on the superposition of \textit{bulge + disk + halo} models. In particular, the total potential consisting of a spherical bulge $\Phi_{\rm b}(R,z)$, an axisymmetric disk $\Phi_{\rm d}(R,z)$ and a spherical dark-matter halo $\Phi_{\rm h}(R,z)$ can be written as follows:
\begin{equation}
\Phi(R,z) = \Phi_{\rm b}(R,z)+ \Phi_{\rm d}(R,z)+ \Phi_{\rm h}(R,z)\,,
\end{equation}
where $R^{2}=x^{2}+y^{2}$ is the Galactocentric distance in polar coordinates and $z$ is the vertical coordinate perpendicular to the disk plane.

Potentials of the bulge and the disk were taken in the form of Miyamoto-Nagai~\cite{Miyamoto1975}, while the dark matter potential is in the form of Navarro-Frenk-White (NFW) profile~\cite{Navarro1997}:
\begin{numcases}{}
\Phi_{\rm b}(R,z)=-\frac{M_{\rm b}}{(r^{2}+b_{\rm b}^{2})^{1/2}},\\
\Phi_{\rm d}(R,z)=-\frac{M_{\rm d}}{\left[R^{2}+\left(a_{\rm d}+\sqrt{z^{2}+b_{\rm d}^{2}}\right)^{2}\right]^{1/2}},\\
\Phi_{\rm h}(R,z)=-\frac{M_{\rm h}}{r}\ln\left(1+\frac{r}{b_{\rm h}}\right)\,,
\end{numcases}
where $\displaystyle r=\sqrt{R^2+z^2}$ is the spherical galactocentric distance, while masses and the scale-lengths of the components are shown in Table~\ref{tab:pot-param} \cite{BajkovaAAT2021,Bajkova2021}.

For the GCs orbital integration we used a high-order parallel dynamical $N$-body code $\varphi$-GRAPE which is based on the fourth-order Hermite integration scheme with hierarchical individual block time steps scheme~\cite{Berczik2011}. More details about the code architecture and special GRAPE hardware can be found in~\cite{Harfst2007}. 

Before moving forward in the analysis of the collisions of the GCs population we have tested our numerical setup in order to keep tracking the GCs whose orbits are the same during backward and forward integration. First, we integrated all $152$ GCs backward for $5$~Gyr then we use the positions of velocities of all the GCs at the end of the simulations and integrate them forward for $5$~Gyr. One could expect that the resulting positions and velocities should be identical to the observed ones. However, we have found that the orbits of 25 GCs are not invertible. These GCs usually pass very close to the galactic center and most likely even an adaptive time-step is not able to capture their motions in close proximity to the centre. Another possibility is a non-integrability of the potential (i.e. it is hard to quantify) and we leave this issue for further studies. These GCs as \textit{to}~(type of orbit) in Table~\ref{tab:GC3} and they were removed from further analysis. As a result,
our final sample consists of $119$ objects.

\bigskip

\begin{minipage}[l]{0.40\textwidth}
\begin{center}
\captionof{table}{Galactic potential parameters.}
\small{\begin{tabular}{lcc}
\hline
\hline
Parameter & Value & Unit \\
\hline
\hline
Bulge mass $M_{\rm b}$  & $1.03\times10^{10}$    & M$_{\odot}$ \\
Disk mass $M_{\rm d}$   & $6.51\times10^{10}$    & M$_{\odot}$  \\
Halo mass $M_{\rm h}$   & $29.00\times10^{10}$   & M$_{\odot}$ \\
Bulge scale param. $b_{\rm b}$   & $0.2672$      & kpc       \\
Disk scale param.  $a_{\rm d}$   & $4.4$         & kpc \\
Disk scale param.  $b_{\rm d}$   & $0.3084$      & kpc  \\
Halo scale param.  $b_{\rm h}$   & $7.7$         & kpc  \\
\hline
\end{tabular}}
\label{tab:pot-param}
\end{center}
\end{minipage}

\section*{\sc gc collision pairs}
\indent \indent 
In order to count the possible maximum number of collisions between all the pairs of GCs we first check all the close encounters during the simulation time with the maximum separation up to $<100$~pc.
Resulting in 2019 and 1973 close encounters in our sample during the backward and forward orbits integration, respectively. 

We define a close encounters as ``collisions'' if (i) the minimum distance between the GCs is less than half the sum of their half-mass radii: 
$dR_{\rm coll} < 2(R_{{\rm hm},i}+R_{{\rm hm},j})$ and (ii) also the relative 
velocity between these objects at the same time $dV_{\rm coll}$ is $<200$~km~s$^{-1}$. 
The first collision condition (i) reduces numbers to only $18$ events, while applying the second (ii) condition we obtained only four reliable collision events.

Fig.~\ref{fig:allcoll} shows the separation parameter as a function of time for backward~(left) and forward~(right) integration where four reliable collisions~(Terzan~3 - NGC~6553, Terzan~3 - NGC~6218, Liller~1 - NGC~6522, Djorg~2 - NGC~6553) are marked by red symbols. It is worth mentioning, that all the colliding GCs were likely originally formed in the MW disk~\cite{Kruijssen2020}. 

\begin{figure*}[htbp!]
\centering
\includegraphics[width=0.4\linewidth]{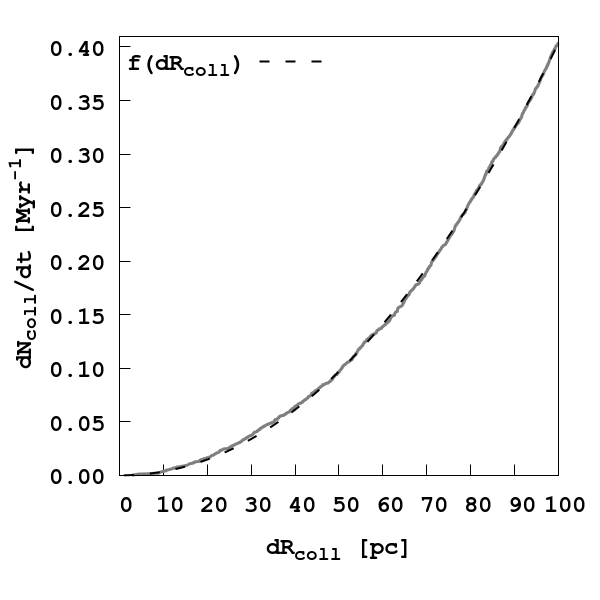}
\includegraphics[width=0.4\linewidth]{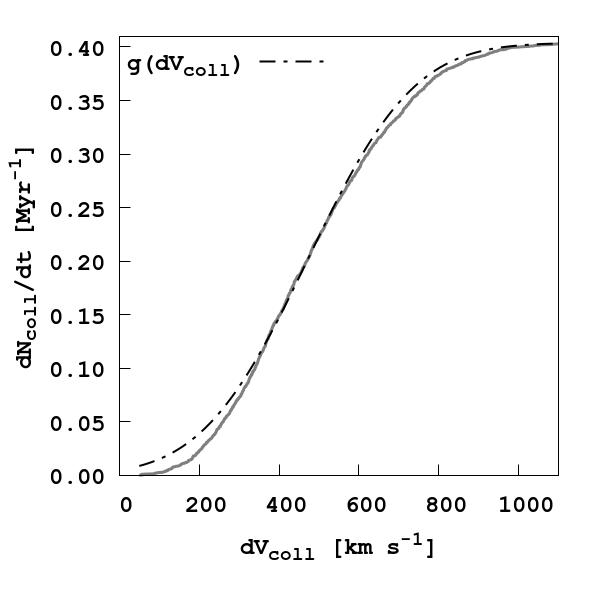}
\caption{GCs collision rate as a function of the relative distance distribution~(left) and relative velocity (right). Black dashed line (left) is a power-law fit $f(x)$ for relative distance (see equation~(\ref{eq:fit})) and dash-dotted line (right) is cumulative distribution function fit $g(x)$ for relative velocity (see equation~(\ref{eq:fitV})).}
\label{fig:stat}
\end{figure*}

In order to estimate the global collision rate, in Fig.~\ref{fig:stat} we show the cumulative collisions number as a function of GCs minimum impact parameter $dR_{\rm coll}$ (left) and relative velocity $dV_{\rm coll}$ (right) at the moment of collision. According to this figure, we can estimate that these plots show that in each ten million years there is at least one collision with the impact parameter less than 50~pc and less than 300~km~s$^{-1}$.

From the cumulative collision number distributions we found the minimum value of impact parameter and relative velocity are $dR_{\rm coll}\approx5$~pc and $dV_{\rm coll}\approx85$~km~s$^{-1}$, respectively. The distribution with the impact parameter was fitted by a simple power-law function:
\begin{equation}
\frac{dN_{\rm coll}}{dt}\Bigl(dR_{\rm coll}\Bigr)=10^{a\cdot\lg(dR_{\rm coll})+b},
\label{eq:fit}
\end{equation}
where the best fit slope parameters are ${\rm a}=2.06$ and ${\rm b}=-4.51$. On the contrary, the velocity distribution was fitted by the cumulative normal distribution function:
\begin{equation}
\frac{dN_{\rm coll}}{dt}\Bigl(dV_{\rm coll}\Bigr)=\frac{1}{2}\Bigl[1+\erf\Bigl(\frac{dV_{\rm coll}-\mu}{\sigma\sqrt{2}}\Bigr)\Bigr],
\label{eq:fitV}
\end{equation}
where we used as a best fit mean value $\mu=472$ and the best fit variance value $\sigma=209$.

\begin{figure}[ht]
\centering
\includegraphics[width=0.99\linewidth]{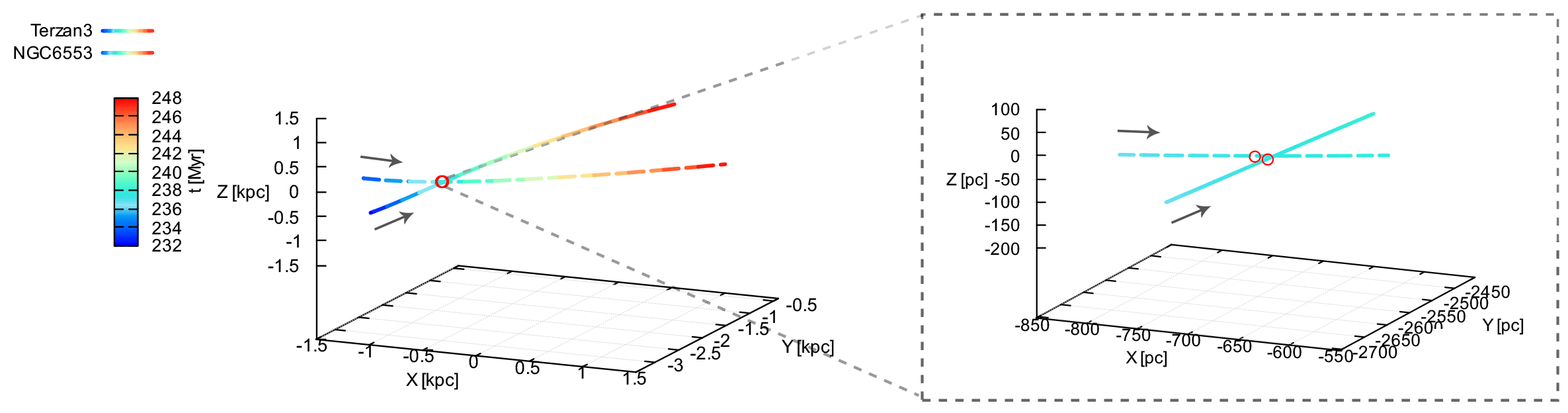}
\includegraphics[width=0.99\linewidth]{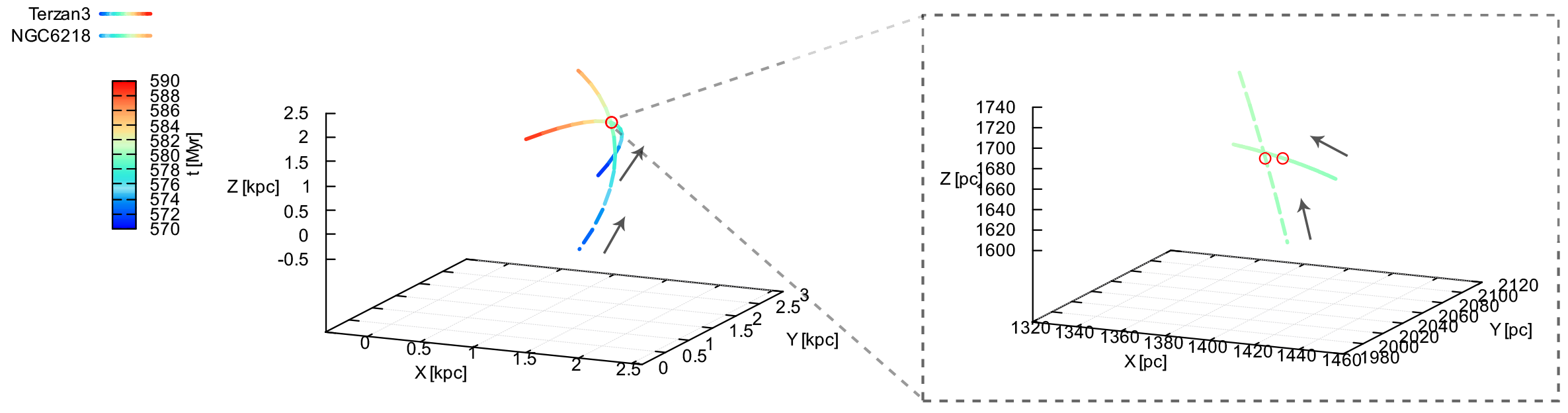}
\includegraphics[width=0.99\linewidth]{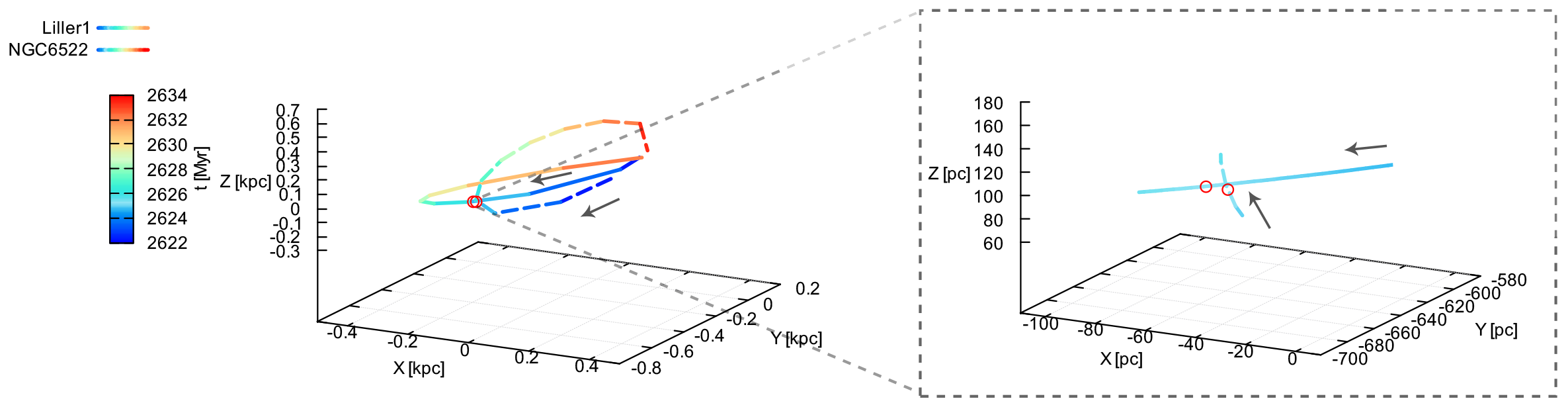}  
\includegraphics[width=0.99\linewidth]{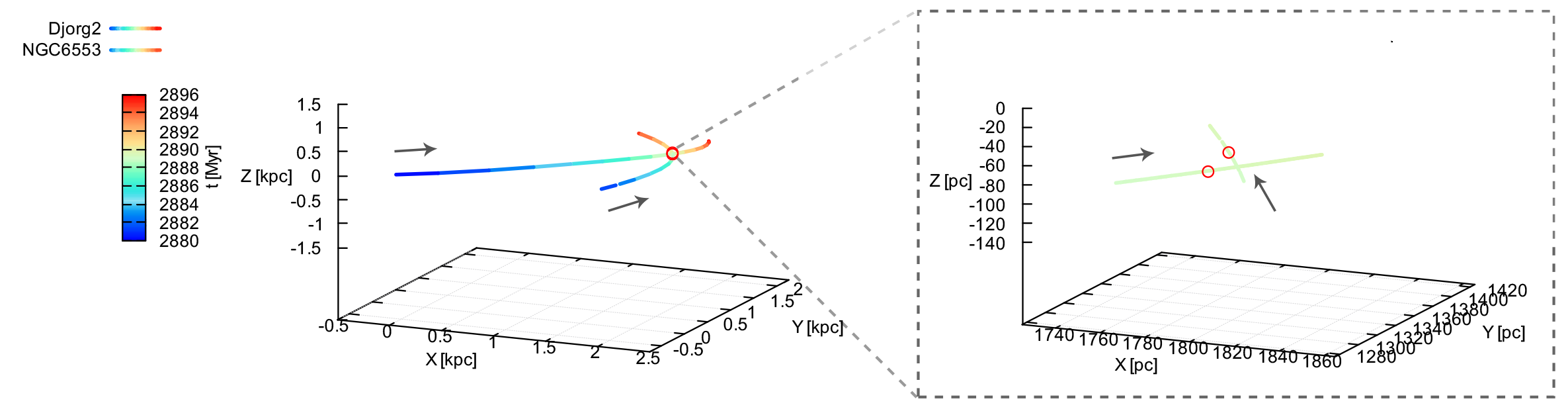} 
\caption{3D orbits of GC “collision” pairs in $\sim20$~Myr (left) and $\sim1$~Myr (right) around collision moment   from (top) to (bottom): (Terzan~3, NGC~6553), (Terzan~3, NGC~6218), (Liller~1, NGC~6522) and (Djorg~2, NGC~6553). Trajectories are colour coded by time, where arrows indicate motion direction and open circles show time moment of collision.}
\label{fig:orbit}
\end{figure}

In Fig.~\ref{fig:orbit} we present the orbits of colliding GCs which are color-coded by time. The time range is about ten million years around the moment of collisions. More detailed the orbital structure is shown in right. The solid line corresponds to the first GC in a pair while the dashed line shows the second one. The intersection of the orbits~(``collision'') is marked as a red circle. Note that our study presents the simplified scenario and more proper study of the GCs ``collisions'' of orbits requires considerations of the gravity interaction between the GCs. 

In Table~\ref{tab:gc-param} we summarize the exact time of ``collisions'' together with the minimum separations and relative velocities at the exact moment of collision. 

\bigskip

\noindent\begin{minipage}[c]{0.40\textwidth}
\begin{center}
\captionof{table}{Characteristics of GC collision pairs.}
\small{
\begin{tabular}{l@{\hspace{0.9\tabcolsep}}l@{\hspace{0.9\tabcolsep}}r@{\hspace{0.9\tabcolsep}}c@{\hspace{0.9\tabcolsep}}r@{\hspace{0.9\tabcolsep}}r@{\hspace{0.9\tabcolsep}}}
\hline
\hline
\multicolumn{1}{c}{GC~1} & \multicolumn{1}{c}{GC~2} & \multicolumn{1}{c}{$dR_{\rm coll}$} & \multicolumn{1}{c}{$dV_{\rm coll}$} & \multicolumn{1}{c}{Time}  & \multicolumn{1}{c}{Prob.}\\
    &      & \multicolumn{1}{c}{(pc)}          & \multicolumn{1}{c}{(km~s$^{-1}$)} & \multicolumn{1}{c}{(Myr)} &  \multicolumn{1}{c}{(\%)}\\
\hline
\hline
Terzan~3  & NGC~6553 & 25.58 & 148.18 &  237 & 22.09 \\
Terzan~3  & NGC~6218 & 10.75 & 183.12 &  580 & 24.32 \\
Liller~1  & NGC~6522 &  9.38 & 185.04 & 2625 & 25.14 \\
Djorg~2   & NGC~6553 & 20.22 & 153.14 & 2889 & 20.23 \\
\hline
\end{tabular}}
\label{tab:gc-param}
\end{center}
\end{minipage}

\bigskip

To check the possible uncertainties introduced by the velocity errors (see Fig.~\ref{fig:PM}) of \textit{Gaia} measurements, we perform extra 10 thousand runs of backward integration with the $\pm~\sigma$ randomly initialized and normally distributed velocities. The $\sigma$ velocity errors (eV$_{\rm R}$, ePMRA and ePMDEC) were taken from the ~\cite{Vasiliev2019} catalog. Following this way we can approximately estimate the probability that our four GSs to collide during the last 5 Gyr of evolution of our Galaxy. From this set of 10 thousand individual runs we obtained that our selected clusters ``collide'' in $\approx$ 11.21~\% of cases. Taking advantage of the randomization in the initial conditions, for each individual GCs pairs we also estimate the lower limit of the collisions probability (see the last column in Table~\ref{tab:gc-param}).

\section*{\sc conclusions}
\indent \indent 
Using the present-day \textit{Gaia} DR2-based catalogs~\cite{Vasiliev2019,Baumgardt2019} we have analyzed the orbits of the Milky Way globular clusters. From 152 GCs we discard 8 objects with large velocity errors and 25 GCs were removed from the analysis due to unstable orbits during backward/forward integration. For the remaining 119 GCs, we analyze both backward and forward orbits calculated in the MW-like external potential using our own developed high order $\varphi$-GRAPE code. Using a complex criteria for the collisions detection we identified four candidate colliding pairs: Terzan~3 - NGC~6553, Terzan~3 - NGC~6218, Liller~1 - NGC~6522, Djorg~2 - NGC~6553. We also estimated the overall collision rate as about one collision with the impact parameter less than 50~pc and less than 300 km s$^{-1}$ per 10~Myr. Our experimental overall close encounter (``collision'') number ($N_{\rm coll}$=4) agrees well with the simple estimation from the collision rate statistical approximations (see Fig.~\ref{fig:stat}).

\section*{\sc acknowledgement}
\indent \indent 
The authors thank the anonymous referee for a very constructive report and suggestions that helped significantly improve the quality of the manuscript. PB and MI acknowledge the support within grant № AP14869395 of the Science Committee of the Ministry of Science and Higher Education of Kazakhstan (``Triune model of Galactic center dynamical evolution on cosmological time scale''). The works of MI and MS were supported by the National Academy
of Sciences of Ukraine under the research project of young scientists № 0121U111799. The work of PB,
MI and MS was also supported by the Volkswagen Foundation under grant № 97778. The work of PB was also supported by the Volkswagen Foundation under a special stipend № 9B870. PB and MI acknowledge the support by the Ministry of Education and Science of Ukraine under the collaborative grant № M/32-23.05.2022.

This work has made use of data from the European Space Agency (ESA) mission GAIA (\url{https://www.cosmos.esa.int/gaia}), 
processed by the GAIA Data Processing and Analysis Consortium (DPAC, \url{https://www.cosmos.esa.int/web/gaia/dpac/consortium}). 
Funding for the DPAC has been provided by national institutions, in particular the institutions participating in the 
GAIA Multilateral Agreement.


\end{multicols}

\begin{table}[ht]
\renewcommand{\arraystretch}{1.0}
\centering
\caption{Initial list of GCs.}
{\small
\begin{tabular}{l@{\hspace{0.95\tabcolsep}}l@{\hspace{0.95\tabcolsep}}c@{\hspace{0.95\tabcolsep}}|l@{\hspace{0.95\tabcolsep}}l@{\hspace{0.95\tabcolsep}}c@{\hspace{0.95\tabcolsep}}|l@{\hspace{0.95\tabcolsep}}l@{\hspace{0.95\tabcolsep}}c@{\hspace{0.95\tabcolsep}}|l@{\hspace{0.95\tabcolsep}}l@{\hspace{0.95\tabcolsep}}c@{\hspace{0.95\tabcolsep}}|l@{\hspace{0.95\tabcolsep}}l@{\hspace{0.95\tabcolsep}}c}
\hline
\hline
ID & Name & Flag & ID & Name & Flag & ID & Name & Flag & ID & Name & Flag & ID & Name & Flag \\
\hline
\hline
1  & NGC~104   &    & 32 & NGC~5634  &    & 63          & NGC~6273  &    & 94  & Terzan~5  & to & 125 & NGC~6656  &    \\
2  & NGC~288   &    & 33 & NGC~5694  &    & 64          & NGC~6284  &    & 95  & NGC~6440  & to & 126 & Pal~8     &    \\
3  & NGC~362   &    & 34 & IC~4499   &    & 65          & NGC~6287  &    & 96  & NGC~6441  &    & 127 & NGC~6681  &    \\
4  & Whiting~1 &    & 35 & NGC~5824  &    & 66          & NGC~6293  & to & 97  & Terzan~6  & to & 128 & NGC~6712  & to \\
5  & NGC~1261  &    & 36 & Pal~5     &    & 67          & NGC~6304  &    & 98  & NGC~6453  &    & 129 & NGC~6715  &    \\
6  & Pal~1     & me & 37 & NGC~5897  &    & 68          & NGC~6316  &    & 99  & NGC~6496  &    & 130 & NGC~6717  & to \\
7  & E~1       & me & 38 & NGC~5904  &    & 69          & NGC~6341  &    & 100 & Terzan~9  & to & 131 & NGC~6723  &    \\
8  & Eridanus  &    & 39 & NGC~5927  &    & 70          & NGC~6325  &    & 101 & Djorg~2   & cc & 132 & NGC~6749  &    \\
9  & Pal~2     &    & 40 & NGC~5946  &    & 71          & NGC~6333  &    & 102 & NGC~6517  & to & 133 & NGC~6752  &    \\
10 & NGC~1851  &    & 41 & BH~176    & me & 72          & NGC~6342  &    & 103 & Terzan~10 &    & 134 & NGC~6760  & me \\
11 & NGC~1904  &    & 42 & NGC~5986  &    & 73          & NGC~6356  &    & 104 & NGC~6522  & cc & 135 & NGC~6779  &    \\
12 & NGC~2298  &    & 43 & FSR~1716  &    & 74          & NGC~6355  &    & 105 & NGC~6535  &    & 136 & Terzan~7  &    \\
13 & NGC~2419  &    & 44 & Pal~14    &    & 75          & NGC~6352  &    & 106 & NGC~6528  &    & 137 & Pal~10    &    \\
14 & Pyxis     &    & 45 & BH~184    &    & 76          & IC~1257   &    & 107 & NGC~6539  &    & 138 & Arp~2     &    \\
15 & NGC~2808  &    & 46 & NGC~6093  &    & 77          & Terzan~2  &    & 108 & NGC~6540  &    & 139 & NGC~6809  &    \\
16 & E~3       &    & 47 & NGC~6121  & to & 78          & NGC~6366  &    & 109 & NGC~6544  & to & 140 & Terzan~8  &    \\
17 & Pal~3     & me & 48 & NGC~6101  &    & 79          & Terzan~4  &    & 110 & NGC~6541  &    & 141 & Pal~11    &    \\
18 & NGC~3201  &    & 49 & NGC~6144  &    & 80          & BH~229    &    & 111 & ESO~280-6 &    & 142 & NGC~6838  &    \\
19 & Pal~4     & me & 50 & NGC~6139  &    & 81$^{\ast}$ & FSR~1758  &    & 112 & NGC~6553  & cc & 143 & NGC~6864  &    \\
20 & Crater    &    & 51 & Terzan~3  & cc & 82          & NGC~6362  &    & 113 & NGC~6558  & to & 144 & NGC~6934  &    \\
21 & NGC~4147  &    & 52 & NGC~6171  &    & 83$^{\ast}$ & Liller~1  & cc & 114 & Pal~7     &    & 145 & NGC~6981  &    \\
22 & NGC~4372  &    & 53 & ESO~452-11 & to & 84          & NGC~6380  & to & 115 & Terzan~12 &    & 146 & NGC~7006  &    \\
23 & Rup~106   &    & 54 & NGC~6205  &    & 85          & Terzan~1  & to & 116 & NGC~6569  &    & 147 & NGC~7078  &    \\
24 & NGC~4590  &    & 55 & NGC~6229  &    & 86          & Ton~2     &    & 117 & BH~261    &    & 148 & NGC~7089  &    \\
25 & NGC~4833  & to & 56 & NGC~6218  & cc & 87          & NGC~6388  &    & 118 & NGC~6584  &    & 149 & NGC~7099  &    \\
26 & NGC~5024  &    & 57 & FSR~1735  & me & 88          & NGC~6402  & to & 119 & NGC~6624  & to & 150 & Pal~12    &    \\
27 & NGC~5053  &    & 58 & NGC~6235  &    & 89          & NGC~6401  &    & 120 & NGC~6626  & to & 151 & Pal~13    &    \\
28 & NGC~5139  &    & 59 & NGC~6254  &    & 90          & NGC~6397  &    & 121 & NGC~6638  & to & 152 & NGC~7492  &    \\
29 & NGC~5272  &    & 60 & NGC~6256  & to & 91          & Pal~6     & to & 122 & NGC~6637  & to &     &           &    \\
30 & NGC~5286  & me & 61 & Pal~15    &    & 92          & NGC~6426  &    & 123 & NGC~6642  & to &     &           &    \\
31 & NGC~5466  &    & 62 & NGC~6266  &    & 93          & Djorg~1   & to & 124 & NGC~6652  & to &     &           &    \\
\hline
\multicolumn{15}{p{.95\textwidth}}{NOTE: Parameters for all GCs was taken from~\cite{Vasiliev2019} with exception for GCs marked~$^{\ast}$ with data from~\cite{Baumgardt2019}. Column Flag contain additional information: me - GC was excluded from the integration due to their significant measurement errors, to - GC was excluded from the integration due to their type of orbit, cc - GC what satisfied “collision” conditions.} \\
\end{tabular}
}
\label{tab:GC3}
\end{table}

\end{document}